\documentclass[aps,prl,twocolumn,a4paper,10pt, longbibliography]{revtex4-2}

\usepackage{bbold}
\usepackage{amsmath,amssymb,mathrsfs}
\usepackage[fleqn]{nccmath}
\usepackage{graphicx}
\usepackage{xcolor}
\usepackage[utf8]{inputenc}

\newcommand{\be}{\begin{equation}}
\newcommand{\ee}{\end{equation}}
\newcommand{\bea}{\begin{eqnarray}}
\newcommand{\eea}{\end{eqnarray}}
\newcommand{\bpm}{\begin{pmatrix}}
\newcommand{\epm}{\end{pmatrix}}

\newcommand{\bs}{\boldsymbol}
\newcommand{\bk}{\mathbf{k}}

\newcommand{\bq}{\mathbf{q}}
\newcommand{\vz}{v_{fz}^0}
\newcommand{\ms}{ \Tilde{m}_y  }
\newcommand{\as}{\Tilde{\alpha}}

\begin{document}

\title{Fermi arcs, Landau levels and magnetic response of the nematic Weyl liquid}

\author{Carlos~Naya}
\affiliation{Institute of Theoretical Physics, Jagiellonian University, Lojasiewicza 11, 30-348 Krak\'ow,Poland}
\affiliation{Department of Physics, Stockholm University, AlbaNova University Center, 106 91, Stockholm, Sweden}
\author{Tomaso~Bertolini}
\affiliation{Department of Physics, Stockholm University, AlbaNova University Center, 106 91, Stockholm, Sweden}
\author{Johan~Carlstr\"om}
\affiliation{Department of Physics, Stockholm University, AlbaNova University Center, 106 91, Stockholm, Sweden}

\begin{abstract}
In classes of Weyl semimetals where the symmetry protects nodes with higher than unit charge, the nematic Weyl liquid appears as interactions destroy this underlying symmetry. In the symmetry-broken phase, the multiple-charge nodes are split into objects of unit charge, the position of which in momentum space is determined by the nematic order parameter. We examine the phenomenology of this phase, focusing on topological edge states and Landau levels. We find that the symmetry-broken phase itself, as well as the orientation of the nematic order are identifiable from the resulting edge states. 
We also find that the nematic order couples to an in-plane magnetic field, indicating that it can be controlled in situ via an external field. Finally, we provide an estimate for the critical coupling where spontaneous symmetry-breaking occurs for contact interaction. 
\end{abstract}

\maketitle

\section{Introduction}
Topological semimetals are characterized by a spectrum that is gapped everywhere, except for a set of gap closing points. Unlike conventional Dirac materials like graphene, the topological nature of these renders them stable against perturbations, and also gives rise to phenomena like the chiral anomaly and surface fermi arcs \cite{Jia2016}. 
Especially the latter of the two has led to an extensive interest in topological materials as building blocks in information technology \cite{Wang2020,Ma2019,Han2018}. Some of these materials also exhibit large magnetoresistance  effects  \cite{Kumar2017}, a property that could be used in magnetic field sensors \cite{Wang2016} and spintronic applications \cite{PhysRevLett.117.146403}.

The prospect of applications based on topological semimetals has also motivated a wide search for materials where gap-closing points are protected by a combination of symmetry and topology, resulting in new types of nodes. Typically, this involves some point-group symmetry and possibly also explicitly broken time-reversal or inversion symmetry. 
Key examples of such systems include band-touching points that carry multiple topological charges \cite{Singh2018,PhysRevLett.108.266802}, involve more than two bands \cite{BeyondDirac, PhysRevLett.119.206402}, or must be classified as line nodes \cite{Bzdusek2016,PhysRevB.96.155105,Bian2016,PhysRevB.93.121113,doi:10.7566/JPSJ.85.013708,doi:10.1063/1.4926545,2017Tanaka}. 

While topologically protected nodal points tend to be very stable against perturbations \cite{PhysRevB.97.161102, PhysRevB.98.241102}, the reliance on symmetry opens up the possibility of spontaneous symmetry-breaking once interactions are taken into account. Such correlation-driven phases can exhibit highly nontrivial topological properties that are tied to the macroscopic order parameter associated with the broken symmetry.
 For example, spontaneous symmetry-breaking in line-node semimetals gives rise to chiral insulators with domain walls that support metallic edge states \cite{PhysRevResearch.5.013069}. 

Coulomb interactions are ubiquitous in electronic systems, and the question of symmetry-breaking is therefore essential to this class of topological systems. To date, there is no comprehensive quantitative data on the threshold required to spontaneously break symmetry in this class of systems, though a few key results exist: From scale transformations it is clear that only linear components of the dispersion survive in the infrared limit \cite{PhysRevB.98.241102}. This is also supported by the finding that the quadratic nodes found in bilayer graphene are unstable against even infinitesimal interaction \cite{PhysRevLett.111.056801}. 
Renormalization group calculations suggest that line-nodes \cite{PhysRevB.96.041113} and Weyl nodes with multiple charges \cite{PhysRevB.95.201102} exhibit a finite threshold against contact interactions, which is consistent with the fact that their dispersion has a linear component. For line-nodes, diagrammatic Monte Carlo calculations provide quantitative estimates in the case of Coulomb interactions \cite{PhysRevResearch.5.013069}. For graphene, Monte Carlo simulations indicate that this material is situated closely to the critical point where it becomes a chiral insulator \cite{PhysRevLett.111.056801}, suggesting that symmetry-breaking is a realistic scenario in real-world electronic systems. 

In this work we focus on the phenomenological properties that emerge when charge-two Weyl nodes are shattered by interactions. This results in a nematic Weyl liquid where the position of the nodal points is tied to the macroscopic order parameter. We do also estimate the critical interaction strength where symmetry-breaking occurs.

\section{Model}
Weyl-nodes with higher than unit charge are obtained as single-charge nodes are effectively stacked in momentum space. This occurs due to discrete rotational symmetries stemming from the lattice structure \cite{PhysRevLett.108.266802}. Crucially, a perturbation which breaks this symmetry will typically split these composite nodes into single charge objects. 
 
 In this work we will consider Weyl semimetals with $C_4$ symmetry, which can host charge-two nodes. 
 In particular, we will focus on surface Fermi arcs and the coupling to an external magnetic field. 
Expanding the dispersion around a nodal point, we obtain an effective low-energy Hamiltonian of the form:
\be \label{H_small_k}
H (\bk)\sim \bpm k_z & (k_x - i k_y)^2 \\( k_x + i k_y)^2 & -k_z \epm,
\ee
which is thus quadratic in the plane, but linear in the $k_z$-direction. 
We will in the following consider a simple two-band tight-binding model given by%
\be \label{ToyH}
H (\bk)\!=\! - 2 (\cos  k_x - \cos k_y)  \sigma_x + 2 \sin k_x \, \sin k_y \, \sigma_y + \sin k_z \, \sigma_z,
\ee
where $\bs \sigma = (\sigma_x, \sigma_y, \sigma_z)$ is the triplet of Pauli matrices. The eigenvalues of the Hamiltonian are given by
\be
\varepsilon_\pm = \pm \sqrt{4 (\cos k_x \!-\! \cos k_y)^2 + 4 \sin^2 k_x \sin^2 k_y + \sin^2 k_z}.
\ee
We thus obtain four nodal points, each corresponding to a double Weyl node, situated at 
\be
\bs q = \{(0,0,0), \, (0,0,\pi), \, (\pi, \pi, 0), \,(\pi, \pi, \pi)\},
\ee
with Chern numbers $\{2,-2,-2,2\}$, respectively. %

If the symmetry which protects the double Weyl node is spontaneously broken, then this will generally result in a nematic order. In the low-energy description, this phase transition is of $U(1)$ universality class, but once higher order terms are included, it will explicitly be broken down to $Z_4$. Once the symmetry is broken, the  dispersion will generally contain perturbations of the form 
\be \label{H_e}
H_{\epsilon} = 2 \epsilon_x \sigma_x + 2 \epsilon_y \sigma_y,
\ee
where $\epsilon$ is a two-component vector that represents the nematic order parameter. The factor 2 has been introduced for convenience.

The perturbed Hamiltonian $H'$ will thus take the form
\bea \nonumber
H' (\bs k) = H(\bs k) + H_\epsilon = - 2(\cos k_x - \cos k_y - \epsilon_x) \sigma_x \\
+ 2 (\sin k_x \sin k_y + \epsilon_y) \sigma_y + \sin k_z \sigma_z,\label{perturbH}
\eea
giving a dispersion 
\bea\nonumber
\varepsilon_\pm = \pm 2 \Big\{(- \cos k_x + \cos k_y + \epsilon_x)^2 \\
+ (\sin k_x \sin k_y + \epsilon_y)^2 + \sin^2 k_z/4\Big\}^{1/2}.
\eea
For a non-vanishing $\varepsilon$, the double Weyl nodes are split into single charge nodes whose dispersion is linear in all directions. Their $q_z$ component does not change ($q_z = 0$ or $\pi$); however, for $q_x$ and $q_y$ we get
\bea\nonumber
q_x= \pm \arccos \left[\frac{1}{2} \left(\epsilon_x \pm \sqrt{4 + \epsilon_x^2 - 4 \sqrt{\epsilon_x^2 + \epsilon_y^2}} \right)\right], \\
q_y\!=\!\mp  \rm sign(\epsilon_y) \arccos \left[\! \frac{1}{2}\! \left(\!-\epsilon_x \pm \sqrt{4 \!+\! \epsilon_x^2\! -\! 4 \sqrt{\epsilon_x^2 \!+\! \epsilon_y^2}} \right) \right]  \nonumber .
\eea

It is convenient to parameterize the symmetry-breaking terms as 
\be 
\epsilon_x = \epsilon \cos \gamma, \qquad \qquad \epsilon_y = \epsilon \sin \gamma,\label{polar}
\ee
where $ \gamma \in (-\pi , \pi]$. The position of the split single nodes then reads
\bea \nonumber
q_x\!=\!  \pm {\rm sign} (\gamma) \arccos \!\left[\!\frac{1}{2}\! \left( \epsilon \cos \gamma \!+\! \sqrt{\epsilon^2 \cos^2 \gamma \!-\! 4 \epsilon+4} \right) \right],\\
 q_y=\mp \arccos \left[ \frac{1}{2} \left(- \epsilon \cos \gamma + \sqrt{\epsilon^2 \cos^2 \gamma -4 \epsilon + 4} \right) \right], \nonumber 
\eea
and
\bea\nonumber
q_x=  \pm \arccos \left[\frac{1}{2} \left( \epsilon \cos \gamma - \sqrt{\epsilon^2 \cos^2 \gamma - 4 \epsilon+4} \right) \right], \\
q_y\!=\!\mp {\rm sign} (\gamma) \arccos \!\left[ \!\frac{1}{2}\! \left(\!-\! \epsilon \cos \gamma \!-\! \sqrt{\epsilon^2 \cos^2 \gamma \!-\!4 \epsilon \!+\! 4} \right) \right] \nonumber \\
 \mp \pi \left[1 - {\rm sign}^2 (\gamma) \right], \nonumber
\eea
where ${\rm sign} (\gamma)$ denotes the {\it signum} function such that ${\rm sign} (0) = 0$.
In particular, we  see that as a function of $\gamma$, the split nodes originating in the double Weyl node at $q_x = 0, q_y=0$ ($q_x = \pi, q_y=\pi$) turn anti-clockwise (clockwise) around this point.

\section{Fermi arcs} \label{Sec_FermiArcs}
Surface states are a hallmark of electronic systems with a nontrivial band structure. In Weyl semimetals, Fermi arcs appear principally as the intermediary between the nontrivial band structure and the trivial vacuum, where the nodes are shifted in momentum space to annihilate. For double nodes, we understand this process to involve the annihilation of two pairs. When the nodes are split, it should be expected that this process is connected to the resulting position of the nodes. The Fermi arcs should therefore be connected to the nematic order.


To obtain solutions for the surface Fermi arcs, we will consider a semi-infinite system in the region $y \leq 0$ with open boundary conditions. Then, we make use of trial functions for an analytical description of surface Fermi arcs \cite{Zhang2016, Ojanen2013, PhysRevResearch.5.013069}. In this description, the systems retains its translation invariance along the $\hat{\bs x}$ and $\hat{\bs z}$ directions, but not in the $\hat{\bs y}$-direction. As a consequence, $k_x$ and $k_z$ are good quantum numbers while $k_y$ is not. Correspondingly, we adopt the substitution $k_y \rightarrow - i \partial_y$. For simplicity, we will consider a pair of Weyl nodes located at $(0,0,0)$ and $(0,0,\pi)$, and expand the Hamiltonian around them at $k_y=0$:
\be \label{Lin_H}
H \sim (2 - 2 \cos k_x - k_y^2) \sigma_x + 2 k_y \sin k_x \sigma_y + \sin k_z \sigma_z.
\ee
This Hamiltonian captures the double nodes at $(0,0,0)$ and $(0,0,\pi)$, the annihilation of which are of primary interest. Notably it does not capture the remaining double nodes, though this is irrelevant for our purposes.

To study the surface states we will consider a trial wave function of the form
\be
\Psi(x,y,z) = \psi_\lambda |x,z \rangle = \begin{pmatrix} \psi_1\\ \psi_2 \end{pmatrix} e^{\lambda y} |x,z\rangle,
\ee
where the dependence on $y$ has been explicitly factorised out from the $|x,z\rangle$ state depending on $x$ and $z$. With $k_y$ no longer being a good quantum number we must substitute $k_y\to -i \partial_y$ giving
\be
H(k_x,k_y,k_z) \rightarrow H(k_x, - i \partial_y, k_z) \rightarrow H(k_x, -i\lambda,k_z).
\ee

Therefore, the problem reduces to the eigenequation
\be
H(k_x,-i\lambda, k_z) \Psi = E \Psi,
\ee
with open boundary conditions at $y=0$. The secular equation, ${\rm det} |H(k_x,-i\lambda, k_z) - E | = 0$, will allow us to obtain the expressions for $\lambda$ as a function of the energy. Since it is a fourth order equation, we find four possible values:
\bea \nonumber
\lambda = \pm \Big[2 \cos k_x ( 1 - \cos k_x) \\
\pm \sqrt{E^2 - \sin^2 k_z - 16 \sin^2 k_x \sin^4 (k_x/2)}\Big]^{1/2}.\label{4lambdas}
\eea

In addition, the eigenequation gives us two different sets of spinors for the eigenvectors, namely,
\be \nonumber
\psi_i = \begin{pmatrix} \sin k_z + E \\ 2 - 2 \cos k_x + \lambda_i^2 + 2 \lambda_i \sin k_x \end{pmatrix}
\ee
and
\be \label{eigenvectors}
\psi_i = \begin{pmatrix} 2 - 2 \cos k_x + \lambda_i^2 - 2 \lambda_i \sin k_x \\ -\sin k_z + E \end{pmatrix}
\ee
where $i=1, 2, 3, 4$ accounts for the different values of $\lambda$. Thus, the general wave function will be the superposition of these states
\be
\psi_\lambda(y,E) = \sum_{i=1}^4 C_i \psi_i e^{\lambda_i y}.
\ee

Imposing open boundary conditions means that
\be
\psi(-\infty,E) = \psi(0,E) = 0.
\ee
The condition at minus infinity implies that only those values of $\lambda$ which are positive (or with a positive real part in case they are complex) are allowed. Thus, just the $\lambda_i$'s with the global plus sign in Eq.(\ref{4lambdas}) are physical. Then, the general wave function becomes
\be
\psi_\lambda(y,E) = C_1 \psi_1 e^{\lambda_2 y} + C_2 \psi_2 e^{\lambda_2 y},
\ee
where 
\bea \nonumber
\lambda_{\substack{1  \\ 2 }}  =&&  \Big[2 \cos k_x ( 1 - \cos k_x) \\
\pm &&\sqrt{E^2 - \sin^2 k_z - 16 \sin^2 k_x \sin^4 (k_x/2)}\Big]^{1/2}.\;\;\; \label{lambdas_E}
\eea

On the other hand, the condition at $y=0$,  leads to
\be
C_1 \psi_1 + C_2 \psi_2 = 0,
\ee
which is a system of two homogenous equations. For nontrivial solutions to exist, it is necessary that
\be
{\rm det} | \psi_1 \quad \psi_2| = 0.
\ee
By imposing it on the two set of eigenvectors in Eq. (\ref{eigenvectors}), we arrive at the  equations
\be
(\sin k_z + E) (\lambda_2 - \lambda_1) (\lambda_1 + \lambda_2 + 2 \sin k_x) = 0,
\ee
\be
(\sin k_z - E) (\lambda_2 - \lambda_1) (\lambda_1 + \lambda_2 - 2 \sin k_x) = 0.
\ee
An obvious solution would be  $\lambda_1 = \lambda_2$, though this corresponds to a double root for $\lambda$, resulting in a trivial wave function. Two additional cases fulfilling these two equations remain however:
\bea
{\rm i)} && \qquad   \lambda_1 + \lambda_2 = 2 \sin k_x, \qquad E = - \sin k_z, \\
{\rm ii)} && \qquad \lambda_1 + \lambda_2 = - 2 \sin k_x, \qquad E = \sin k_z.
\eea
Recalling the fact that $\lambda_i > 0$ (or $\Re(\lambda_i) > 0$), it is clear that $\lambda_1 + \lambda_2 >0$. Therefore, we can conclude that 
\be
\left\{ \begin{array}{lll} \lambda_1 + \lambda_2 = 2 \sin k_x, & \quad E = - \sin k_z, & \quad {\rm if} \: k_x > 0, \\
 \lambda_1 + \lambda_2 = - 2 \sin k_x, & \quad E = \sin k_z, & \quad {\rm if} \: k_x < 0, 
 \end{array} \right.
\ee
which can be also written as
\be \label{E_surface}
\lambda_1 + \lambda_2 = 2 |\sin k_x|, \qquad E_{\rm surf} = - {\rm sign}(k_x) \sin k_z,
\ee
where $E_{\rm surf}$ denotes the energy dispersion of the surface states. 

If we insert this energy into the expression for $\lambda_1$ and $\lambda_2$, we obtain,
\be
\lambda_{\substack{1  \\ 2 }}  = \left[ 2 \cos k_x (1-\cos k_x) \pm 4 i |\sin k_x| \sin^2 (k_x/2)\right]^{1/2},
\ee
which implies that  
\be \label{product_lambdas}
\lambda_1 \lambda_2 = 4 \sin^2(k_x/2).
\ee
Taking this into consideration and multiplying Eq. (\ref{E_surface}) by $\lambda_1$, we arrive at
\be
\lambda_1^2 - 2 |\sin k_x| \lambda_1 + 4 \sin^2 (k_x/2) = 0.
\ee
A similar expression can be found for $\lambda_2$, providing the following form of the parameters $\lambda$:
\be
\lambda_{\substack{1  \\ 2 }} = |\sin k_x| \pm 2 i \sin^2(k_x/2).
\ee
On the other hand, choosing one set of spinors from Eq. (\ref{eigenvectors}) allows us to write the wave function of the surface states,
\be
\Psi(\bs x) = C \begin{pmatrix} H(-k_x) \\ H(k_x) \end{pmatrix} (e^{\lambda_1 y} - e^{\lambda_2 y}) |x,z\rangle,
\ee
where $H(k_x)$ is the Heaviside step function.

\begin{figure}[t]
\begin{center}
\includegraphics[width=\linewidth]{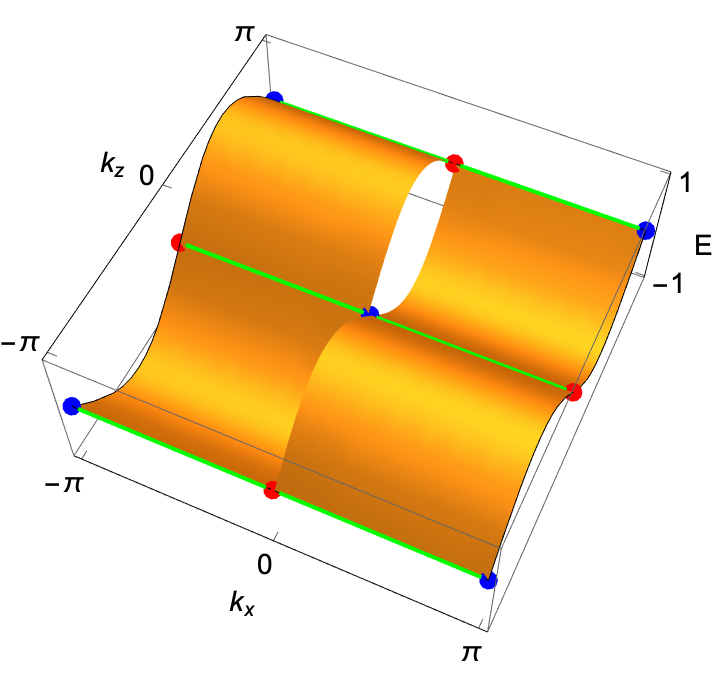}
\end{center}
\caption{Energy dispersion of the surface states together with the projection of the Weyl nodes and Fermi arcs. Blue (red) balls correspond to the projection of Weyl nodes with positive (negative) chirality. Green lines connecting them represent the Fermi arcs. For a better visualisation, equivalent Weyl nodes and Fermi arcs at $k_{x, \, z} - 2 \pi$ are also shown.}
\label{Fig_ESurf}
\end{figure}

From Eq. (\ref{E_surface}) it is clear the energy dispersion shows a discontinuity at $k_x = 0$ (and the same happens at $k_x = \pi$). This indicates we have two distinguishable surface states: one for $-\pi < k_x < 0$, and another one when $0 < k_x < \pi$, as depicted in Fig. \ref{Fig_ESurf}. We have also found four Fermi arcs, which correspond to the surface states at the Fermi level $E_{\rm F} = 0$. They are the straight lines $k_z = 0$ and $k_z=\pi$  connecting the Weyl nodes at $k_x = 0$ and $k_x = \pm \pi$ (the nodes at $k_x = \pi$ and $k_x = -\pi$ are equivalent). The fact that there are two Fermi arcs connecting each pair of nodes results from the topological charge being two.

It should be noted that, despite considering the linearised Hamiltonian in $k_y$, Eq. (\ref{Lin_H}), the situation is equivalent to the original Hamiltonian (\ref{ToyH}). This is due to the fact that, when considering the projection of the Weyl nodes onto the XZ plane, the split nodes from the linearised Hamiltonian sit on top of each other and become indistinguishable, so the total Chern number is still two.

We now consider how the surface states change when introducing a symmetry-breaking term of the form given in Eq. (\ref{H_e}). In this case, the projection of the Weyl nodes onto the surface does not coincide, and all Chern numbers are one. As before, we consider a semi-infinite system $y \leq 0$, so the component of the momentum $k_y$ is no longer a good quantum number.  For simplicity, we will again employ an  expansion in $k_y$ of the Hamiltonian as introduced in Eq. (\ref{Lin_H}). Including the symmetry-breaking term, it reads
\be
H \!=\! (2 \!-\! 2 \cos k_x \!-\!k_y^2 + 2 \epsilon_x) \sigma_x \!+\! 2 (k_y \sin k_x \!+\! \epsilon_y ) \sigma_y \!+\! \sin k_z \sigma_z.
\ee

In this linearized perturbed Hamiltonian, the positions of the split nodes are slightly modified, and their analytical form becomes more involved. For this reason, we do not include it here, though it can be found by solving the system of equations
\be
k_y \sin k_x + \epsilon_y =0, \qquad \qquad 2-2 \cos k_x - k_y^2 + 2 \epsilon_x = 0.
\ee

As before, the possible values for $\lambda$ as function of the energy are given by the secular equation, which reduces to solve
\bea \nonumber
E^2 \!=\! \lambda^4\!+\!4 (\cos^2 k_x \!-\!  \cos k_x \!+\! \epsilon \cos \gamma) \lambda^2 \!-\! 8 i \epsilon \sin \gamma \sin k_x \lambda \\
 + 4 (1 - \cos k_x) (1-\cos k_x + 2 \epsilon \cos \gamma) \nonumber 
  + 4 \epsilon^2 + \sin^2 k_z   ,\\
 \label{perturbed_lambdas}
\eea
where we use the polar form of the symmetry-breaking term given in Eq. (\ref{polar}), with $\gamma$ representing the angle, and $\epsilon$ the magnitude. Although it is a quartic equation in $\lambda$, only two solutions will fullfil the requirement $\Re(\lambda) > 0$, which we denote as $\lambda_1$ and $\lambda_2$. Then, a general solution would be a combination of these cases, but we recall that to have a nontrivial solution the corresponding spinors $\psi_1$ and $\psi_2$ must satisfy ${\rm det} | \psi_1 \quad \psi_2| = 0$. Again, we have two sets of spinors which translates the latter condition into the same two equations as in the unperturbed case, resulting in the same expressions as in Eq. (\ref{E_surface}).

It is important to note that the current expressions for $\lambda_1$ and $\lambda_2$ obtained from Eq. (\ref{perturbed_lambdas}) do not fulfil this condition everywhere. In fact, this is a welcomed feature since they obey it just between the projection of the Weyl nodes onto the surface, defining in this way the Fermi arcs as the lines $k_z =0, \pi$ between nodes. 

\begin{figure}[]
\begin{center}
\includegraphics[width=\linewidth]{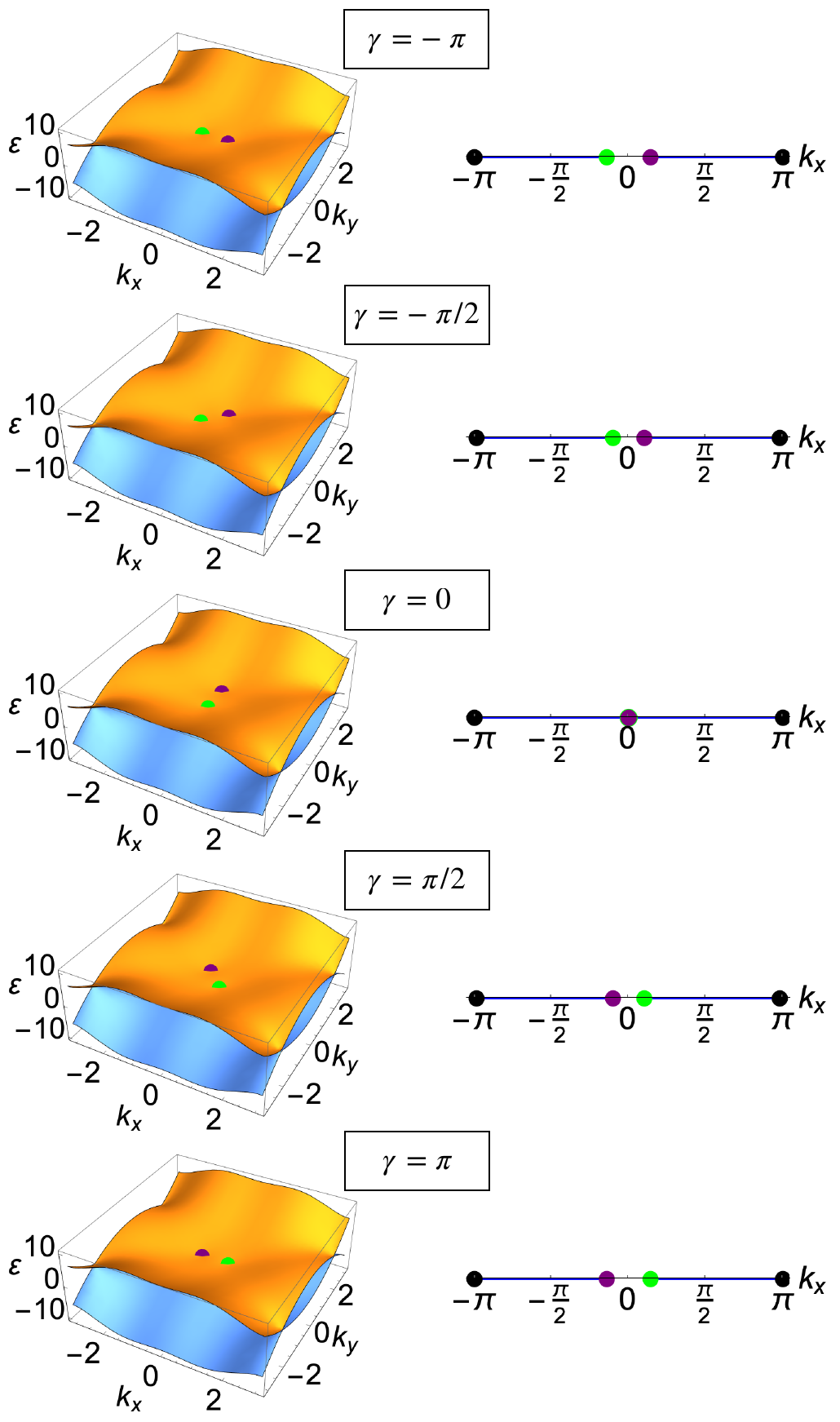}
\end{center}
\caption{{\it Left}: Energy bands for the perturbed system on the $k_z = 0$ plane for $\epsilon = 0.1$ and different values of $\gamma$. We highlight the split nodes around the origin with dots of different colors to easily track their motion in the Brillouin zone for a varying $\gamma$. {\it Right}: Fermi arcs connecting the Weyl nodes for $\epsilon = 0.1$ and varying $\gamma$. The same criterion for the split Weyl nodes has been used as in the left plots. For completeness, we also show the other two Weyl nodes as black balls.}
\label{Fig_Nodes-Gamma}
\end{figure}

Considering the effect of changing the polar angle $\gamma$ from $-\pi$ to $\pi$, we see that the position of the split Weyl nodes approximately describe a semi-circle until they are interchanged (see Fig. \ref{Fig_Nodes-Gamma} left). In particular, focusing on the projection of the nodes which originate from the double node at $k_x = k_y = 0$, we see that the distance between them is larger when $\gamma = \pm \pi$, while they coincide for $\gamma=0$. However, as shown in Fig. \ref{Fig_Nodes-Gamma} right, we find a discontinuity in the Fermi arcs. For $\gamma >0$ they do not pass each other but interchange the nodes at their endpoints. 

\section{External magnetic field and Landau levels} \label{Sec_LandauLevels}

In this section, we will consider the magnetic response of a double Weyl node and the closely related split node which occurs when the system becomes nematic. We will initially consider the symmetric case defined by Eq. (\ref{ToyH}), and expand around the node to lowest order, giving a low-energy Hamiltonian %
\be \label{H_expansion}
H= \chi_x (k_x^2 - k_y^2) \sigma_x + 2 k_x k_y \sigma_y +\chi_z k_z \sigma_z
\ee
where we have introduced
\be
\chi_x = \cos q_x = \pm 1, \qquad \chi_z = \cos q_z = \pm 1,
\ee
so that the chirality $\chi$ of the node is defined by $\chi = \chi_x \chi_z$.

To couple the system to an external field we consider the displacement in momentum $\bs k$ by the vector potential $\bs A$, namely,
\be
\bs k \longrightarrow \bs k' = \bs k + \bs A.
\ee
We will initially consider a magnetic field along the $\bs{ \hat z}$ direction. Below, we will also consider both out-of-plane ($\bs{\hat z}$) and in-plane ($\bs{\hat x}$) fields in a nematic system.
For the out-of-plane case, $\bs B = B \bs{\hat z}$, we will work with the axial gauge
\be
\bs A = (- B y/2, B x/2, 0),
\ee
which allows us to define the ladder operators as functions of the new momenta as follows
\be \label{a_Bz}
a = \frac{k_x'- i k_y'}{\sqrt{2 B}}, \qquad a^\dagger = \frac{k_x'+ i k_y'}{\sqrt{2 B}}.
\ee
Hence, the Hamiltonian reduces to
\be
H = \chi_x B (a^2 + (a^\dagger)^2) \sigma_x + i B (a^2 - (a^\dagger)^2) \sigma_y + \chi_z k_z \sigma_z.
\ee	
Now, to find the Landau levels we need to solve the eigenequation $H | \Phi \rangle = E | \Phi \rangle$. Writing the eigenstate $| \Phi \rangle$ as $ | \Phi \rangle = \left( | \phi_1 \rangle, | \phi_2 \rangle \right)^{\rm T}$, we are left with the system of equations
\be
\chi_z k_z | \phi_1 \rangle + B \left[ (\chi_x + 1) a^2 + (\chi_x -1) (a^\dagger)^2 \right] | \phi_2 \rangle = E | \phi_1 \rangle,
\ee
\be
B \left[ (\chi_x - 1) a^2 + (\chi_x +1) (a^\dagger)^2 \right] | \phi_1 \rangle - \chi_z k_z | \phi_2 \rangle = E |\phi_2 \rangle.
\ee

Focusing on the case  $\chi_x = 1$, which corresponds to the nodes located at $q_x = q_y = 0$, the system becomes
\be \label{SystemEq1}
\chi_z k_z | \phi_1 \rangle + 2 B a^2 | \phi_2 \rangle = E | \phi_1 \rangle,
\ee
\be \label{SystemEq2}
2 B (a^\dagger)^2 | \phi_1 \rangle - \chi_z k_z | \phi_2 \rangle = E | \phi_2 \rangle.
\ee
Then, taking the first equation and plugging it into the second, we arrive at
\be
\left[ 4 B^2 a^\dagger a (a^\dagger a-1) + k_z^2 \right] |\phi_2 \rangle = E^2 |\phi_2 \rangle.
\ee
This is an eigenvalue equation similar to the harmonic oscillator which results in the energy levels
\be
E_n = \pm \sqrt{4 B^2 n (n-1)+k_z^2},
\ee
with eigenstates
\be
|\phi_1 \rangle = c \frac{2 B \sqrt{n (n-1)}}{E_n - \chi_z k_z} \, |n-2\rangle, \qquad \qquad |\phi_2 \rangle = c |n\rangle,
\ee
where $c$ is a normalisation constant. This expression is valid for $n$ being an integer greater than 2. This is due to the fact that for $n = 0, 1$,  $| \phi_1 \rangle$ vanishes, and we get instead
\be
E_{0,1} = - \chi_z k_z,
\ee
\be
|\Phi_0 \rangle = (0, |0\rangle)^{\rm T}, \qquad \qquad |\Phi_1 \rangle = (0, |1\rangle)^{\rm T}.
\ee

On the other hand, if we consider the nodes at $q_x = q_y = \pi$, i.e., $\chi_x = - 1$, we obtain similar expressions for the energy 
\be
E_{0,1} = \chi_z k_z, \qquad \qquad E_n = \pm \sqrt{4 B^2 n (n-1)+k_z^2},  
\ee
and eigenstates
\be
|\Phi_0 \rangle = (|0\rangle, 0)^{\rm T}, \qquad \qquad |\Phi_1 \rangle = (|1\rangle, 0)^{\rm T},
\ee

\be
|\phi_1 \rangle = c |n\rangle,  \qquad \qquad |\phi_2 \rangle= - c \frac{2 B \sqrt{n (n-1)}}{E_n + \chi_z k_z} \, |n-2\rangle, .
\ee
Notably, the sign of $E_{0,1}$ changes, reflecting the fact that these energy levels are chiral. The chirality of the Weyl node is given by $\chi = \chi_x \chi_z = \pm 1$:
\bea
E_{0,1} \!=\! - \chi k_z, \;E_n \!=\! \pm \sqrt{4 B^2 n (n\!-\!1)\!+\!k_z^2}, \;{\rm for} \; n \geq 2.\;\;
\eea

\begin{figure*}[htb]
\begin{center}
\includegraphics[width=\textwidth]{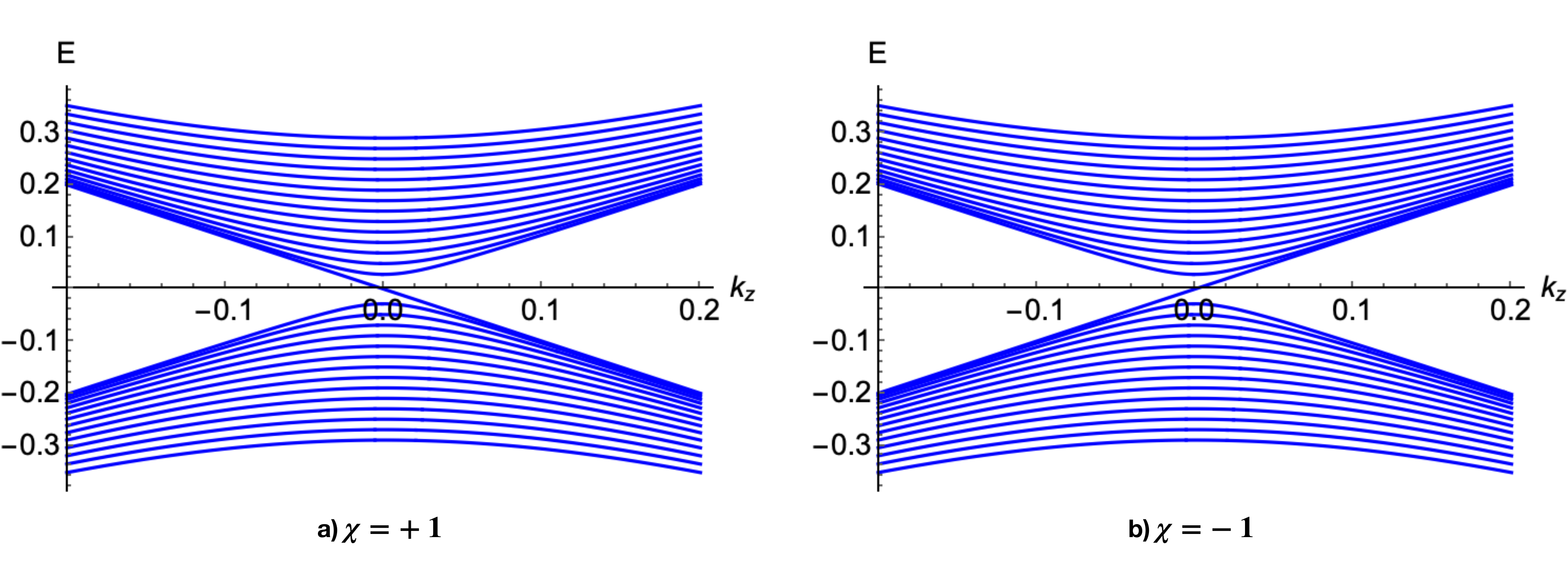}
\end{center}
\caption{Landau levels for an out of plane magnetic field $\bs B = 10^{-2} \bs{\hat z}$. As for single Weyl nodes, we have chiral states depending on a) positive or b) negative chirality. The higher character of our critical points (double Weyl nodes) is manifest in the fact that these particular states are doubled degenerated in energy.}
\label{Fig_LL-OutOfPlane}
\end{figure*}

As can be seen in Fig. \ref{Fig_LL-OutOfPlane}, the Landau levels associated with the double Weyl nodes do not differ qualitatively from those of a single nodes except for the fact that they are doubly degenerate, reflecting the higher charge of the node. As a result, we obtain the pair of chiral states $E_0,\;E_1$. 

Next, we consider the perturbed system with split double nodes. Due to the complexity of the resulting system, we will consider the Hamiltonian Eq. (\ref{H_expansion}) up to second order in momentum $\bs k$. After introducing the perturbation given by Eq. (\ref{H_e}), we obtain
\be
H = \left[ \chi_x (k_x^2 - k_y^2) + 2 \epsilon_x \right] \sigma_x +2 (k_x k_y + \epsilon_y) \sigma_y + \chi_z k_z \sigma_z.
\ee
Hence, the split nodes are located at
\be
(q_x, q_y) = \sqrt \epsilon \left( \pm {\rm sign} (\gamma) \sqrt{1 - \cos \gamma}, \mp \sqrt{1+\cos \gamma} \right)
\ee
if $\chi_x = 1$, and
\be
(q_x, q_y) = \sqrt \epsilon \left( \pm \sqrt{1+\cos \gamma} \right), \mp {\rm sign} (\gamma) \sqrt{1 - \cos \gamma})
\ee
if $\chi_x = -1$,
while their position along the $z$ axis remains unchanged.

By expanding this Hamiltonian around a single node, we can see it takes the form $H = H_0 + H_1$, where $H_0$ is the Hamiltonian of a single node and $H_1$ appears as a warping term, quadratic in $\bs k$:
\be \label{H0}
H_0 = \begin{pmatrix} \chi_z k_z &  (\chi_x k_x - i k_y)\Gamma \sqrt{\frac{2}{B}}\\
 (\chi_x k_x + i k_y) \Gamma^\dagger\sqrt{\frac{2}{B}} & - \chi_z k_z \end{pmatrix},
\ee
with $\Gamma=\sqrt{2B}(q_x - i \chi_x q_y)$ and
\be \label{H1}
H_1 = \begin{pmatrix} 0 & \chi_x (k_x - i \chi_x k_y)^2 \\
 \chi_x (k_x + i \chi_x k_y)^2 & 0 \end{pmatrix}.
\ee
To solve this problem, we will treat $H_1$ as a perturbation. We will consider both an out-of plane and an in-plane field. 

Starting with an out-of-plane magnetic field, $\bs B = B \bs{\hat z}$, and using the creation and annihilation operators of Eq. (\ref{a_Bz}), we obtain
%
\be
H_0\! =\!\! \begin{pmatrix} \chi_z k_z &  \!\Gamma  \!\left[\chi_x(a\!+\!a^\dagger) \!+\! a \!-\! a^\dagger \right] \\
 \Gamma^\dagger\left[\chi_x(a\!+\!a^\dagger) \!-\! a \!+\! a^\dagger \right]& \!-\! \chi_z k_z \end{pmatrix}.
\ee
The resulting eigensystem $H_0 \Phi = E \Phi$ to solve is similar to the unperturbed case, namely,
\bea\nonumber
 E | \phi_1 \rangle=\chi_z k_z |\phi_1 \rangle + 2 (q_x - i \chi_z q_y) \sqrt \frac{B}{2} \\
\times[ (1 + \chi_x) a - (1 - \chi_x) a^\dagger ] | \phi_2 \rangle 
\eea
and
\bea\nonumber
 E | \phi_2 \rangle= - \chi_z k_z | \phi_2 \rangle 2 (q_x + i \chi_z q_y) \sqrt \frac{B}{2} \\
\times[ (1 + \chi_x) a^\dagger - (1 - \chi_x) a ] | \phi_1 \rangle.
\eea
 Applying the same approach as for Eq. (\ref{SystemEq1}) and (\ref{SystemEq2}), namely inserting one equation into the other, we obtain the Landau levels 
\be
E_0 = - \chi k_z, \; E_{n, \pm} = \pm \sqrt{k_z^2 + 16 \epsilon B n}, \;{\rm for} \; n \geq 1,
\ee
with eigenstates
\be
\Phi_0 = \begin{pmatrix} 0 \\ | 0 \rangle \end{pmatrix},\;\; \Phi_{n,\pm} = c \begin{pmatrix} \frac{2 \sqrt{2 B} (q_x - iq_y)}{E_{n,\pm} - \chi_z k_z} \sqrt n |n -1\rangle \\ |n\rangle \end{pmatrix},
\ee
for $\chi_x = 1$, and
\be
\Phi_0 = \begin{pmatrix} | 0 \rangle \\ 0 \end{pmatrix}, \;\;\Phi_{n,\pm} = c \begin{pmatrix} |n\rangle \\ - \frac{2 \sqrt{2 B} (q_x + iq_y)}{E_{n,\pm} + \chi_z k_z} \sqrt n |n -1\rangle \end{pmatrix},   \;
\ee
for $ \chi_x = -1$, where $c$ is a normalization constant.

Next, we need to consider the contribution from $H_1$. Using ladder operators, we obtain \\$H_1 =$
\be
\begin{pmatrix} \!0\! &  \!\frac{\chi_x B}{2} \!\left[\! (1\!+ \!\chi_x) a \!+\! (\!1\! -\! \chi_x\!) a^\dagger \right]^2 \\  \frac{\chi_x B}{2}\! \left[ \!(\!1\!-\! \chi_x\!) a \!+\! (\!1\! +\! \chi_x\!) a^\dagger \! \right]^2\! & 0\! \end{pmatrix}.\;
\ee
Treating this perturbatively, the first order contribution, $E^{(1)}$, vanishes while at second order, $E^{(2)}$, we obtain
\bea
E^{(2)}_0 = 0,\; E^{(2)}_{1, \pm} = \mp \frac{4 B^2}{\sqrt{16 \epsilon B+k_z^2}},\\
 E^{(2)}_{n,\pm} = \pm \frac{4 B^2 n^2}{\sqrt{16 \epsilon B n + k_z^2}}, \; {\rm for} \; n\geq 2. 
\eea
Interestingly, we see that the chiral state $n=0$ does not receive any correction and does not couple to the nematic order parameter $\epsilon$. The energy levels do not depend on the angle $\gamma$, as should be expected  from symmetry.

Finally, we can  consider the most interesting case of an in-plane magnetic field. For the low-energy description we can without loss of generality set $\bs B = B \bs{\hat x}$. In this scenario, symmetry permits that the energy does couple to the phase of the nematic angle, $\gamma$.
 The starting point is again the Hamiltonian $H = H_0 + H_1$ given by Eq. (\ref{H0}) and (\ref{H1}). However for convenience, when solving the resulting eigenequation once the system is coupled to an external magnetic field, we will introduce a unitary transformation given by the SU(2) matrix
\be \label{Vmatrix}
V = \frac{1}{\sqrt 2} \begin{pmatrix} 1 & 1 \\ - e^{i \delta} & e^{i \delta} \end{pmatrix},
\ee
where $\delta$ is defined from the node position as follows:
\be \label{Def-delta}
q_x + i q_y = q \, e^{i \delta}.
\ee
Applying this transformation, the leading Hamiltonian $H_0$ reads
\be
\tilde{H_0} = V^\dagger H_0 V = \begin{pmatrix} - 2 \chi_x \sqrt{2 \epsilon} k_x & \chi_z k_z - 2 i \sqrt{2 \epsilon} k_y \\ \chi_z k_x + 2 i \sqrt{2 \epsilon} k_y & 2 \chi_x \sqrt{2 \epsilon} k_x \end{pmatrix}.
\ee

Then, the momentum $k_x$ appears in the diagonal elements only, which makes the system easier to solve once we introduce the ladder operators, and $k_x$ is the only surviving component of the momentum. 

\begin{figure*}[htb]
\begin{center}
\includegraphics[width=\textwidth]{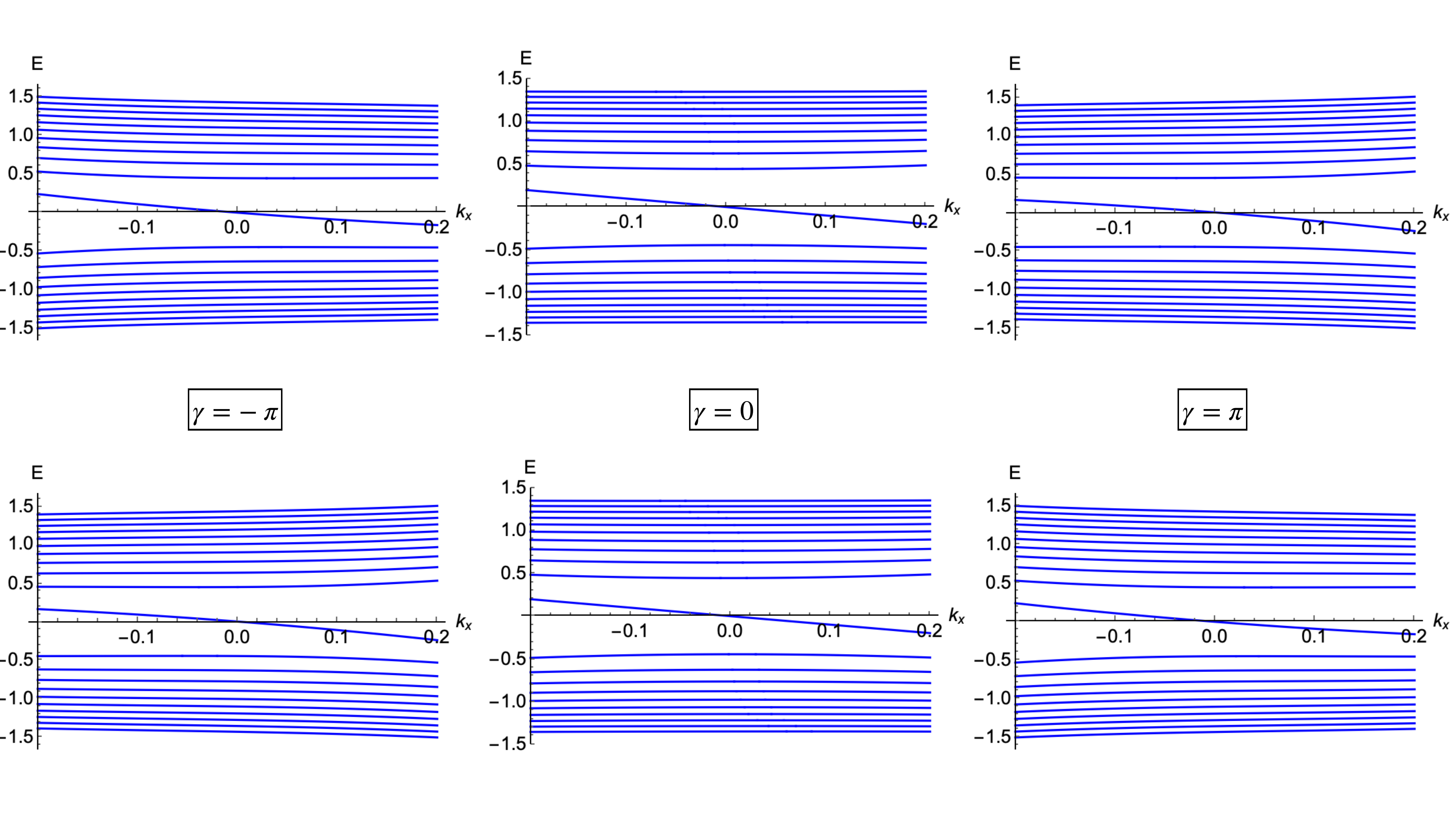}
\end{center}
\caption{Landau levels of a pair of split nodes with $\chi_x = 1$ up to second order in perturbation theory for an in plane magnetic field $\bs B = 10^{-2} \bs{\hat x}$ for different values of $\gamma$.}
\label{Fig_LL-InPlane-2ndOrder}
\end{figure*}

Introducing creation and annihilation operators of the form 
\be \label{a_Bx}
a = \frac{k_y'- i k_z'}{\sqrt{2 B}}, \qquad a^\dagger = \frac{k_y'+ i k_z'}{\sqrt{2 B}},
\ee
the eigensystem $\tilde{H_0} \Phi = E \Phi$ takes the form
\begin{widetext}
\be
\begin{pmatrix} - 2 \chi_x \sqrt{2 \epsilon} k_x & i \chi_z \sqrt \frac{B}{2} (a - a^\dagger) - 2 i \sqrt{\epsilon B} ( a + a^\dagger) \\ i \chi_z \sqrt \frac{B}{2} (a - a^\dagger) + 2 i \sqrt{\epsilon B} ( a + a^\dagger) & 2 \chi_x \sqrt{2 \epsilon} k_x \end{pmatrix} \; \begin{pmatrix} | \phi_1 \rangle \\ |\phi_2 \rangle \end{pmatrix} = E \begin{pmatrix} | \phi_1 \rangle \\ | \phi_2 \rangle \end{pmatrix}.
\ee
For convenience, we will consider the particular value $\epsilon = 1/8$, then the system reduces to
\be
\begin{pmatrix} - \chi_x k_x & - i (1 - \chi_z ) \sqrt \frac{B}{2} a - i(1 + \chi_z) \sqrt \frac{B}{2} a^\dagger \\ 
i (1 + \chi_z) \sqrt \frac{B}{2} a + i (1 - \chi_z) \sqrt \frac{B}{2} a^\dagger & \chi_x k_x \end{pmatrix} \begin{pmatrix} | \phi_1 \rangle \\ | \phi_2 \rangle \end{pmatrix} = E \begin{pmatrix} | \phi_1 \rangle \\ | \phi_2 \rangle \end{pmatrix}.
\ee
%
Now, we can consider the cases $\chi_z = \pm 1$, obtaining a system of equations which we can solve by the same approach as for the out-of-plane field case. Hence, we arrive at
\be \label{LL-perturb-Bx}
E_0 = - \chi k_x, \; E_{n,\pm} = \pm \sqrt{k_x^2 + 2 B n},\; {\rm for} \; n \geq 1,
\ee
while the eigenstates are given by
\be
\Phi_ 0 = \begin{pmatrix} |0 \rangle \\ 0 \end{pmatrix}, \;\Phi_n = c \begin{pmatrix} | n \rangle \\ \frac{i \sqrt{2 B n}}{E_{n,\pm} - \chi_x k_x} |n - 1 \rangle \end{pmatrix}, \;{\rm for} \; \chi_z = 1,\;\;\;\;
\Phi_ 0 \!=\! \begin{pmatrix} 0 \\ |0 \rangle\end{pmatrix}, \;\Phi_n \!=\! c \begin{pmatrix} - \frac{ i \sqrt{2 B n}}{E_{n,\pm} + \chi_x k_x} |n\! -\! 1 \rangle \\ | n \rangle \end{pmatrix}\!, \; {\rm for} \; \chi_z \!=\! -\!1,
\ee%
where $c$ is a normalization constant. In Fig. \ref{Fig_LL-InPlane-0thOrder} the energy levels for a split node of each chirality can be seen, revealing no dependende on the nematic order parameter at this level of the expansion.
\begin{figure*}[htb]
\begin{center}
\includegraphics[width=\linewidth]{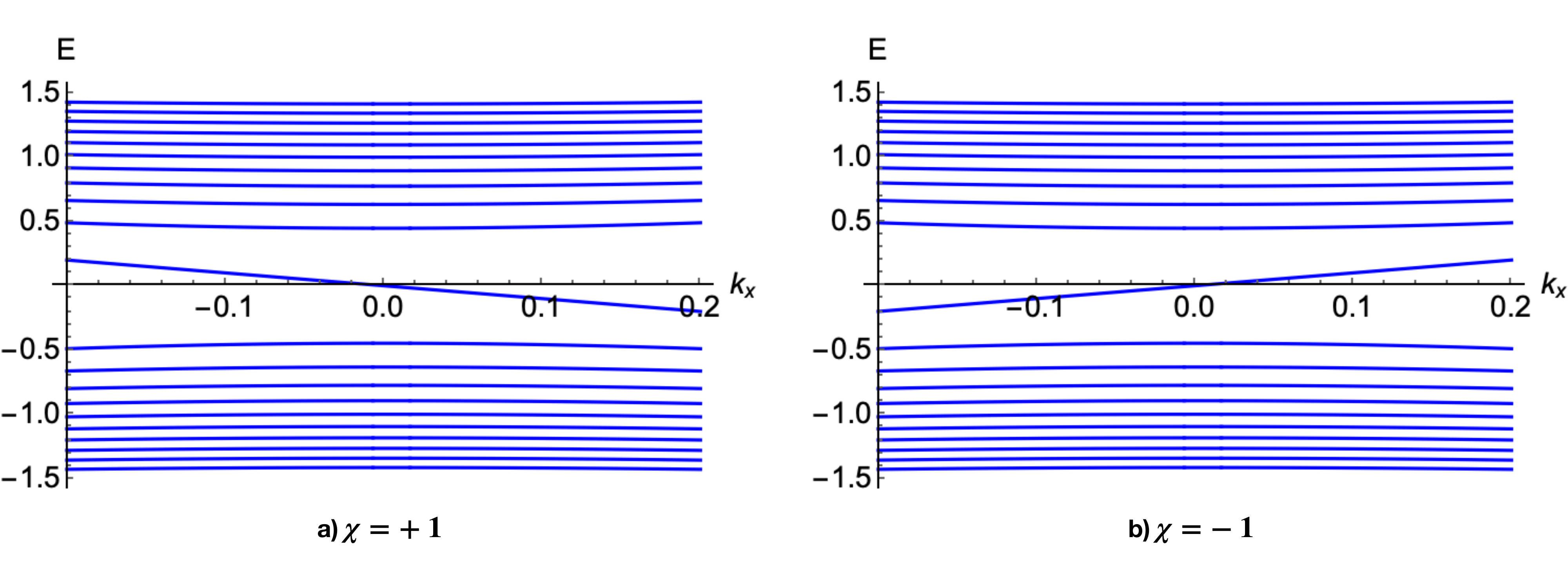}
\end{center}
\caption{Leading contribution to the Landau levels of a split node for an in plane magnetic field $\bs B = 10^{-2} \bs{\hat x}$ once a perturbation is included: a), positive; b), negative chirality. It should be noted that, at this zeroth order in perturbation theory, the nematic order parameter do not play any role.}
\label{Fig_LL-InPlane-0thOrder}
\end{figure*}

To take into account the contribution from $H_1$, we apply the unitary transformation  from Eq. (\ref{Vmatrix}), obtaining,
%
\be
\tilde{H_1} = V^\dagger H_1 V = \begin{pmatrix} -\chi_x \cos \delta (k_x^2 - k_y^2) - 2 \sin \delta k_x k_y & i \chi_x \sin \delta (k_x^2 - k_y^2) - 2 i \cos \delta k_x k_y \\
- i \chi_x \sin \delta (k_x^2 - k_y^2) + 2 i \cos \delta k_x k_y & \chi_x \cos \delta (k_x^2 - k_y^2) + 2 \sin \delta k_x k_y \end{pmatrix}.
\ee
\end{widetext}
After introducing the creation and annihilation operators, we can calculate the corresponding perturbative corrections to the energy levels Eq. (\ref{LL-perturb-Bx}). In particular, we see that the first order corrections do not vanish in this case, taking the form
\bea
E^{(1)}_0 = -\chi \cos \delta \left( k_x^2 - \frac{B}{2} \right), \;\;\\
 E^{(1)}_{n,\pm} \!=\! \frac{\cos \delta}{2 E^{(0)}_{n,\pm}} (2 k_x^3 \!+\! 2 k_x B n\!+\!\chi E^{(0)}_{n,\pm} B ), \; {\rm for} \; n \geq 1.\;\;
\eea
%
%
Notably, this contribution is odd under the reflection $\{q_x,q_y\}\to -\{q_x,q_y\}$ since in this case $\delta\to\delta+\pi$ according to Eq. (\ref{Def-delta}). This implies, that when adding the energy contribution to a pair of nodes, situated at $\bq,-\bq$, the net contribution vanishes. Specifically, the contributions are
\be
\cos \delta = \left \{ \begin{array}{ll} \pm \sin(\gamma/2), & \qquad {\rm if} \; \chi_x = 1, \\
\pm \cos(\gamma/2), & \qquad {\rm if} \; \chi_x = -1, \end{array} \right.\label{sign}
\ee
where the sign depends on the specific node being considered. To obtain a net energy contribution we have to go to second order. Here, we find 
%
\begin{widetext}
\bea
E^{(2)}_0= &&\left(\sin^2 \delta - \frac{3}{4} \cos^2 \delta \right) \chi k_x B,\\
E^{(2)}_{1, \pm} =&& - \frac{B \sin^2 \delta}{2 E^{(0)}_{1,\pm}} \left( \chi k_x E^{(0)}_{1,\pm} + 7 k_x^2 + 6 B \right) + \frac{3 \chi k_x B^2 \sin^2 \delta}{E^{(0)}_{1,\pm} (E^{(0)}_{1,\pm} - \chi k_x)} \nonumber \\
 &&+ \frac{B \cos^2 \delta}{4 E^{(0)}_{1,\pm}} (5 k_x^2 + B) - \frac{3 \chi k_x B^2 \cos^2 \delta}{2 E^{(0)}_{1,\pm} (E^{(0)}_{1,\pm}-\chi k_x)} + \frac{B^3 \cos^2 \delta}{(E^{(0)}_{1,\pm})^3},\\
E^{(2)}_{2,\pm} =&&  - \frac{B \sin^2 \delta}{E^{(0)}_{2,\pm}} (5 k_x^2 + 12 B) + \frac{4 \chi k_x B^2 \sin^2 \delta}{E^{(0)}_{2,\pm} (E^{(0)}_{2,\pm}-\chi k_x)} + \frac{B \cos^2 \delta}{8 E^{(0)}_{2,\pm}} (3 \chi k_x E^{(0)}_{2,\pm} + 17 k_x^2 + 8 B) \nonumber \\
 &&- \frac{9 \chi k_x B^2 \cos^2 \delta}{2 E^{(0)}_{2,\pm} (E^{(0)}_{2,\pm}-\chi k_x)} + \frac{8 B^3 \cos^2 \delta}{(E^{(0)}_{2,\pm})^3},\\
E^{(2)}_{n,\pm} =&&  - \frac{B \sin^2 \delta}{E^{(0)}_{n,\pm}} \left[ (2 n+1) k_x^2 + 3 n^2 B \right] + \frac{2 n \chi k_x B^2 \sin^2 \delta}{ E^{(0)}_{n,\pm} (E^{(0)}_{n,\pm} - \chi k_x)} + \frac{B \cos^2 \delta}{4 E^{(0)}_{n,\pm}} \left[ (2n + 3) k_x^2 + n^2 B \right] \nonumber \\
&& - \frac{3 n \chi k_x B^2 \cos^2 \delta}{2 E^{(0)}_{n,\pm} (E^{(0)}_{n,\pm} - \chi k_x)} + \frac{n^3 B^3 \cos^2 \delta}{(E^{(0)}_{n,\pm})^3}, \qquad {\rm for} \; n \geq 3.
\eea
\end{widetext}
The essential feature of this expression is that it contains terms of the form $\sin^2 \delta$ and $\cos^2 \delta$, which are even under the reflection $\bq\to-\bq$. Consequently, the total energy of the two nodes depends on the angle $\delta$, of the nematic order. The low-energy Hamiltonian (Eq. \ref{H0}, \ref{H1}) is invariant under a simultaneous rotation of the magnetic field direction and the nematic order around the $z$-axis, and so the implication of this result is that the nematic order couples to the magnetic field direction. Correspondingly, the order parameter can be controlled by a magnetic field. The landau levels for $\chi_x = 1$ up to second order in perturbation theory are displayed in Fig. \ref{Fig_LL-InPlane-2ndOrder}.

\section{Stability of the double node} \label{stability}

Renormalization group calculations suggest that double Weyl nodes become unstable at a finite interaction threshold for contact interaction \cite{PhysRevB.95.201102}. To establish an estimate for this threshold we employ Fock theory, since the Fock term typically contains most of the corrections to the dispersion both in the symmetric \cite{PhysRevB.98.241102} and symmetry-broken phases \cite{PhysRevResearch.5.013069} for semimetals. 
 
We consider a low-energy description of a double Weyl node with bare Fermi velocities $v_f^0 , v_{fz}^0$:
\bea
H_0(\bk)=v_f^0  (k_x^2-k_y^2 )\sigma_x + v_f^0 (k_x k_y)\sigma_y  +v_{fz}^0 k_z \sigma_z  . \label{H0}
\eea
We take the interaction to be $V(\bk)= \alpha $. In this scenario, a self-consistent treatment of Fock theory can be conducted semi-analytically to produce the symmetry-breaking term. 
The self-energy thus satisfies 
\begin{align} \label{selfcE}
\Sigma(\omega_m,\bk) &= - \frac{1}{\beta} \sum\limits_{n} \int \frac{d^3q}{(2\pi)^3}   \frac{\alpha}{G_0^{-1}(\omega'_n,\bq) - \Sigma(\omega'_n,\bq) } \\
G_0(\omega_n, \bq) &= \frac{1}{i \omega_n - H_0 (\bq)} 
\end{align} 
The self energy $\Sigma$ becomes independent of frequency and momentum since the interaction is local in time and space. The resulting self-energy then reduces to constant symmetry-breaking terms of the form $\Sigma = m_x \sigma_x +m_y \sigma_y +m_z \sigma_z$. The term proportional to $m_z$, if present, would only shift the node along the $z$-axis and not alter the nature of the bands so it will be neglected in the following.  
In the low-temperature limit the Matsubara frequency can be integrated over to obtain
\begin{widetext}
\bea\nonumber
\Sigma &&= \frac{\pi \alpha}{(2 \pi)^4}  \int  d^3 q \, \frac{H_0(q)+\Sigma}{|H_0(q) +\Sigma|} \\ \nonumber
    &&=  \frac{\pi \alpha}{(2 \pi)^4}  \int d^3 q  \frac{v_f^0(q_x^2-q_y^2) \sigma_x + m_x \sigma_x + v_f^0 q_x q_y \sigma_y + m_y \sigma_y    }{\sqrt{( (v_f^0 q_x^2 -v_f^0 q_y^2 +m_x   )^2 +(v_f^0 q_x q_y + m_y)^2 +\vz q_z^2   )}} \\
    &&=  \frac{\pi \alpha}{(2 \pi)^4} \int d^3 q \, \frac{ m_x \sigma_x +  m_y \sigma_y    }{\sqrt{ (v_f^0 q_x^2 -v_f^0 q_y^2 +m_x   )^2 +(v_f^0 q_x q_y + m_y)^2 +(\vz)^2 q_z^2   }}\label{aaa}
\eea
\end{widetext}
where we note that odd terms in the nominator vanish under integration.
For symmetry reasons, we can without loss of generality take the nematic order parameter to be oriented along the $y-$axis, setting $m_x=0$.

We work on a cylindrical domain with radius and height $\Lambda$. It is convenient to introduce the dimensionless momentum variable \textbf{p} according to
\be
\bq = \Lambda \mathbf{p}
\ee
In the new variables Eq. (\ref{aaa}) reads
\be
\Sigma \!=\! \frac{\pi}{(2 \pi)^4}\! \frac{\alpha \Lambda}{v_f^0}\! \int \!d^3 \!q  \frac{ m_y \sigma_y    }{\sqrt{ ( q_x^2 \!-\!q_y^2)^2\! +\!(q_x q_y \!+\! \frac{m_y}{\Lambda^2 v_f^0} )^2 \!+\! \frac{(\vz)^2 q_z^2}{ {v_f^0}^2 \Lambda^2}   }}
\ee
It is useful to define the rescaled parameters 
\be
\ms = \frac{m_y}{v_f^0 \Lambda^2} , \quad \as =  \frac{\pi}{(2 \pi)^4} \frac{\alpha \Lambda}{v_f^0}, \quad  \Tilde{v}_z = \frac{\vz }{ v_f^0 \Lambda}.
\ee
To obtain a numerical solution, we set $\Tilde{v}_z$ to unity, giving a self consistency equation for $\ms$ on the form
\bea \label{aab}
\ms &&= \as \ms  \int d^3p \, \frac{1}{\sqrt{ (p_x^2-p_y^2)^2 + (p_x p_y +\ms)^2 + p_z^2  }} \\
 &&= \as \ms I(\ms) .
\eea
 Equation (\ref{aab}) can be numerically solved to predict the onset of a nematic order at a finite interaction strength, and this critical value can be analyticaly found to be $\as_c = 1/ I(0) \approx 0.11 $.

\begin{figure}[!htb]
	\includegraphics[width=\linewidth]{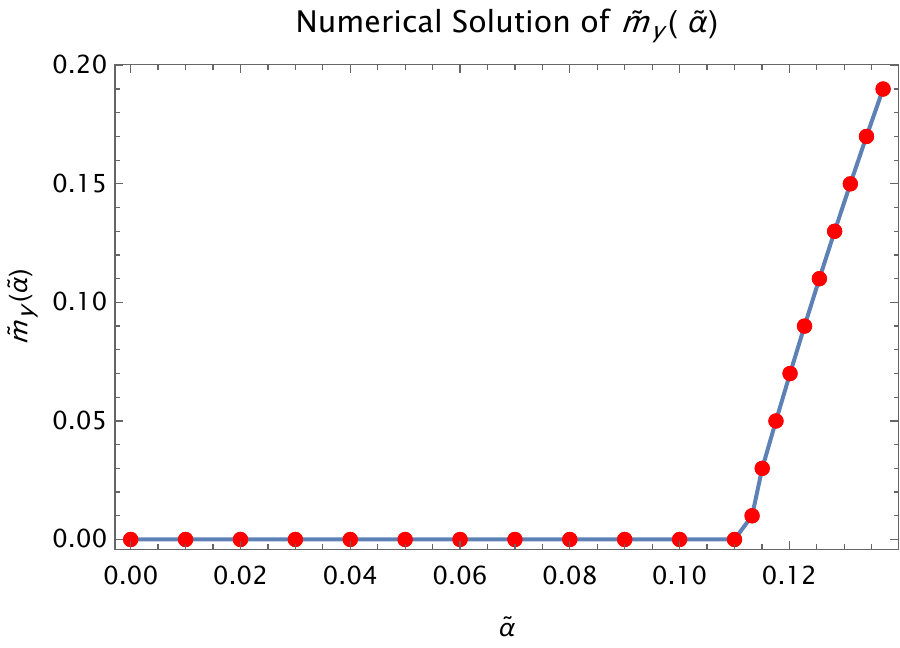}
	\caption{ 		{\bf Numerical self-consistent solution for the gap} with contact interaction, as a function of the rescaled coupling strength $\as$. Solutions to eq. (\ref{aab}) are plotted as red dots, the solid line is a guide to the eyes.}
	\label{doubenodecritical}
\end{figure}

\section{Conclusions}
We have examined topological edge states and the magnetic response of a nematic Weyl liquid. We find that the symmetry-broken state has clear experimental indicators in the form of topological edge states, and that these are coupled to the nematic order parameter, which thus becomes measurable. We also find that for an in-plane magnetic field, the nematic order parameter couples to the Landau levels. This implies that the orientation of the nematic order can be controlled by an external magnetic field, even allowing nematic vortices to be created for instance. 
The coupling between an external magnetic field and the emerging order parameter is also observed in symmetry-broken line-node semimetals \cite{PhysRevResearch.5.013069}, suggesting that this feature may be universal in such systems. 
Finally, we estimate the critical coupling for contact interaction, finding this value to be finite, in agreement with renormalization group calculations \cite{PhysRevB.95.201102}. 

This work was supported by the Swedish Research Council (VR) through grant 2018-03882 and Stiftelsen Olle Engkvist via grant 204-0185. C.N. is further supported by the Polish National Science Centre (NCN), Grant No. NCN 2020/39/B/ST2/01553. J. C. would like to thank Lars Fritz for important input and discussions.

\bibliography{biblio.bib}

\end{document}